\documentclass[11pt,hyper,a4paper]{JHEP}
   
\usepackage{epsfig}
\usepackage{latexsym}
\usepackage{graphicx}
\usepackage{amsmath}
\usepackage{amsfonts}   % if you want the fonts
\usepackage{amssymb}    % if you want extra symbols
\usepackage{float}
\usepackage{bm}
\def\ee{e^+e^-\to h^0Z}

\def\lsim{\raise0.3ex\hbox{$\;<$\kern-0.75em\raise-1.1ex\hbox{$\sim\;$}}}
\def\gsim{\raise0.3ex\hbox{$\;>$\kern-0.75em\raise-1.1ex\hbox{$\sim\;$}}}
\def    \be            {\begin{equation}}
\def    \ee            {\end{equation}}
\def    \bea           {\begin{eqnarray}}
\def    \eea           {\end{eqnarray}}

\def\sw2{sin^2 \theta_w}

\def\a^tau{\alpha_{\tau}}

\def\beq{\begin{equation}}
\def\eeq{\end{equation}}
\def\beqa{\begin{eqnarray}}
\def\eeqa{\end{eqnarray}}

\newcommand{\newc}{\newcommand}
\newc\BR{BR}
\newc{\akappa}{A_{\kappa} }

\newc\deltagmtwo{\delta (g-2)_{\mu}} 
\newc\deltaamu{\Delta a_{\mu}}

\def\anti{\overline}

\newc{\haa}{BR\(h_1\to a_1 a_1\)}
\newc{\abb}{BR\(a_1\to b\anti{b}\)}
\newc{\hbb}{BR\(h_1\to b\anti{b}\)}
\newc{\abund}{\Omega h^2}

\newc\bsgamma{b\rightarrow s \gamma }
\newc\bxsgamma{\overline{B}\rightarrow X_{s}\gamma}
\newc\brbsgamma{\BR(\overline{B}\rightarrow X_s\gamma)}

\title{The $\mu\nu$SSM with an extra $U(1)$}
\author{}
 \author{Javier Fidalgo\\
         Departamento de F\'{\i}sica Te\'{o}rica UAM
         and Instituto de F\'{\i}sica Te\'{o}rica UAM/CSIC,\\
         Universidad Aut\'{o}noma de Madrid (UAM), Cantoblanco,
         28049 Madrid, Spain\\
         E-mail: \email{javier.fidalgo@uam.es}}
 \author{Carlos Mu\~noz\\
         Departamento de F\'{\i}sica Te\'{o}rica UAM
         and Instituto de F\'{\i}sica Te\'{o}rica UAM/CSIC,\\
         Universidad Aut\'{o}noma de Madrid (UAM), Cantoblanco,
         28049 Madrid, Spain\\
         E-mail: \email{carlos.munnoz@uam.es}}

\abstract{\small
The $\mu\nu$SSM solves the $\mu$ problem of the MSSM and generates correct neutrino masses by simply using 
right-handed neutrinos. This mechanism implies that only dimensionless trilinear terms, breaking $R$-parity, are present in the superpotential.
We present an extension of the $\mu\nu$SSM with an extra $U(1)$ gauge symmetry. We use the extra $U(1)$ charges of the matter fields to forbid the presence in the superpotential of renormalizable and non-renormalizable
baryon number violating operators, the trilinear operator producing a domain wall problem, and bilinear operators such as the $\mu$ term and the Majorana masses.
We apply the anomaly cancellation conditions associated to the extra $U(1)$, to constrain the values
of the $U(1)$ charges. We find that six assignments of the $U(1)$ charges to the matter fields
are viable, once extra matter is introduced.
In particular, three generations of vector-like color triplets and $SU(2)_L$ doublets, as well as six Standard Model singlets are necessary. 
Electroweak symmetry breaking is viable in the model, with wide regions of the parameter space fulfilling the experimental constraints on the existence of a new gauge boson $Z'$.
Neutrinos and the extra gaugino mix with the MSSM neutralinos, producing a generalized see-saw matrix that can reproduce the experimental results on neutrino masses. Finally, we have estimated the tree-level upper bound on the lightest Higgs mass, finding that it can be as large as about 120 GeV. \\
}


\preprint{
\rightline{FTUAM-11-58, IFT-UAM/CSIC-11-74, November 2011}
}

\begin{document}

\section{Introduction}

The $\mu$ from $\nu$ Supersymmetric Standard Model 
($\mu\nu$SSM) \cite{Primer paper munu,Unpublished notes,MuNuSSMreview} is defined by the following superpotential:
\begin{align}\label{superpotential munuSSM}
W &=\epsilon_{ab} \left(
Y_u^{ij} \, \hat H_2^b\, \hat Q^a_i \, \hat u_j^c +
Y_d^{ij} \, \hat H_1^a\, \hat Q^b_i \, \hat d_j^c +
Y_e^{ij} \, \hat H_1^a\, \hat L^b_i \, \hat e_j^c +
Y_\nu^{ij} \, \hat H_2^b\, \hat L^a_i \, \hat \nu^c_j 
\right)
\nonumber\\
&-\epsilon_{ab} \lambda^{i} \, \hat \nu^c_i\,\hat H_1^a \hat H_2^b
+
\frac{1}{3}
\kappa^{ijk} 
\hat \nu^c_i\hat \nu^c_j\hat \nu^c_k\,,
%\label{superpotential}}
\end{align}
where we take $\hat H_1^T=(\hat H_1^0, \hat H_1^-)$, $\hat H_2^T=(\hat
H_2^+, \hat H_2^0)$, 
$\hat Q_i^T=(\hat u_i, \hat d_i)$, 
$\hat L_i^T=(\hat \nu_i, \hat e_i)$, $i,j,k=1,2,3$ and $a,b=1,2$ are generation and
$SU(2)$ indices, respectively, and $\epsilon_{12}=1$.
This superpotential contains 
%The superpotential of the $\mu\nu$SSM 
%contains, 
in addition to the usual
%usual 
Yukawas for quarks and charged leptons,
Yukawas for neutrinos
$Y_{\nu} \hat H_2\,  \hat L \, \hat \nu^c$, terms of the type
$\lambda \hat \nu^c \hat H_1\hat H_2$ producing an
effective  $\mu$ term through right-handed sneutrino vacuum expectation values (VEVs) of the order of the
electroweak (EW) scale, $\mu\equiv
\lambda \langle \tilde \nu^c \rangle$,
and terms of the type $\kappa \hat \nu^c \hat \nu^c \hat \nu^c$  
avoiding the existence of a Goldstone boson and producing an EW-scale see-saw through the generation of
effective Majorana masses
$\kappa \langle \tilde \nu^c \rangle$.

Thus the 
$\mu\nu$SSM
solves the
$\mu$-problem \cite{muproblem} of the Minimal Supersymmetric Standard Model (MSSM) \cite{mssm}
and generates light neutrino masses
by simply using right-handed neutrino superfields.
% is an alternative to the Minimal Supersymmetric Standard Model (MSSM) \cite{mssm}
Note that the above terms in the superpotential produce the explicit breaking
of $R$-parity in this model.
The size of the breaking can be 
%The breaking of R-parity in the $\mu\nu$SSM can be 
easily understood realizing that in the limit where 
%neutrino Yukawa couplings 
$Y_{\nu}$ are vanishing, the 
$\hat \nu^c$ are 
ordinary singlet superfields like the $\hat S$ 
of the Next-to-Minimal Supersymmetric Standard Model (NMSSM) \cite{ana}, 
without any connection with neutrinos,
and 
%{\bf this model would be like the
%NMSSM 
%(although having three singlets instead of one) 
%with three singlets},
%where 
$R$-parity is therefore conserved.
Once $Y_{\nu}$ are switched on, 
the 
%singlets
$\hat \nu^c$ become right-handed neutrinos, and, as a consequence, $R$-parity
is broken. 
% {\bf This breaking has to be small because} of the
% EW-scale seesaw implying that {\bf the values of $Y_{\nu}$ 
%neutrino Yukawa couplings 
% cannot be larger than $10^{-6}$} (like the electron Yukawa) to reproduce the correct neutrino masses ($\lsim 10^{-2}$ eV).
Thus the breaking is small because
the EW-scale see-saw implies small values for $Y_{\nu}\sim 10^{-6}$.

Since the $\mu\nu$SSM is a very well motivated and attractive model, several phenomenological studies have been carried out.
In \cite{Segundo paper munu,MuNu SCPV}, the parameter space, the spectrum and the vacua of the $\mu\nu$SSM were analyzed in detail.
The neutrino sector was studied in \cite{Segundo paper munu,MuNu SCPV,MuNu Indian,1loop corrections indios MuNu}, obtaining that current neutrino data (the measured mass differences and mixing angles) can be easily reproduced.
Analyses of possible signals at colliders were also carried out. Since the Lightest Supersymmetric Particle (LSP) is no longer stable due to the breaking of $R$-parity, not all supersymmetric chains must yield missing energy events.
In \cite{MuNu Indian,MuNu Hirsch,Porod} the decays of the lightest neutralino were discussed, as well as the correlations of the decay branching ratios with the neutrino mixing angles.
Let us remark that the breaking of $R$-parity generates a peculiar structure for the mass matrices, and this has to be taken into account in the computations mentioned above.
In particular, the presence of right and left-handed sneutrino VEVs leads to mixing 
of neutralinos with left- and right-handed neutrinos, and as a consequence a generalized matrix of the see-saw type. Besides, there is also the mixing of the neutral Higgses with the sneutrinos producing 8$\times$8
neutral scalar mass matrices, and this extended Higgs sector could be very helpful for testing the $\mu \nu$SSM at colliders \cite{MuNu Hirsch,indios3}.
In \cite{MuNu Higgs sector colliders}, special emphasis was put in the decays of the Higgses and viable benchmark points for LHC searches were provided.

Concerning cosmological issues, dark matter and baryon asymmetry have been analyzed in the model.
% Since the lightest supersymmetric particle (LSP) is not stable when $R$-parity is broken,
%In this context
% the neutralino \cite{reviewmio}
%\cite{reviewmio} 
% or the right-handed sneutrino \cite{sneutrino}, 
%\cite{sneutrino},
% with very short lifetimes,
% are no longer candidates for the dark matter of the Universe.
The
gravitino, present in the local supersymmetric version of the model, 
could be a good dark matter candidate as discussed 
in \cite{MuNu Gravitino DM,MuNu Gravitino DM 2}, where 
its possible detection through the observation of a monochromatic gamma-ray line in the \textit{Fermi} satellite was also studied.
In \cite{MuNu Baryogenesis}, the generation of the baryon asymmetry of the Universe was analyzed in detail in the context of the $\mu\nu$SSM, with the interesting result that EW baryogenesis can be realized.

Once $R$-parity is not a symmetry of the model, 
lepton and baryon number violating terms in the superpotential
like 
\begin{equation}\label{baryon}
\epsilon_{ab} \left(\lambda'''_{ijk} \hat L_i^a \hat L_j^b \hat e^c_k 
+ 
\lambda'_{ijk} \hat L_i^a \hat Q_j^b \hat d^c_k 
+ 
\mu_i  \hat L_i^a\hat H_2^b \right)\,\,\ , \,\,\,\,
\lambda''_{ijk} \hat d^c_i \hat d^c_j \hat u^c_k\ ,
\end{equation}
are in principle allowed by gauge
invariance. As it is well known, to avoid too fast proton decay mediated by
the exchange of squarks of masses of the order of the EW scale,
the presence together of terms of the type  
$\hat L \hat Q \hat d^c$ and 
$\hat d^c \hat d^c \hat u^c$ must be forbidden, unless
we impose very stringent bounds such as e.g.
%$\lambda'^*_{112}\dot \lambda''_{112} \lsim 2\times 10^{-27}$.
$\lambda'^*_{112} \lambda''_{112} \lsim 2\times 10^{-27}$.
Clearly, these values for the couplings are not very natural, and
for constructing viable supersymmetric models one usually 
forbids at least one of the operators
$LQd^c$ or $d^c d^c u^c$. 
The other type of operators above are not so stringently suppressed, and
therefore still a lot of freedom remains \cite{dreiner}.

% One possibility to avoid the problem of proton decay in the MSSM is to impose
% $R$-parity conservation (+1 for particles and -1 for superpartners).
% Actually this forbids all the four operators above and thus protects the proton.
%Nevertheless, the choice of $R$-parity is {\it ad hoc}. 

There are several ways to avoid this problem. One is to assume 
that there are other discrete 
symmetries, 
like e.g. baryon triality which only forbids the baryon violating operators \cite{dreiner3}.
%Obviously, for all these symmetries R-parity is violated.
Another one comes from string constructions, where the matter superfields can be located in different sectors or have different extra $U(1)$ charges, in such a way that 
some operators violating $R$-parity can be forbidden \cite{old}, 
but others can be allowed.

Another problem is related to the absence of a $\mu$ term as well as 
Majorana masses for neutrinos in the superpotential (\ref{superpotential munuSSM}), since both type of bilinear terms are in principle allowed by gauge invariance.
As for the proton decay problem above, we have several solutions.
The fact that only dimensionless trilinear terms are present in the superpotential of the $\mu\nu$SSM,
can be explained invoking a $Z_3$ symmetry, 
as it is usually done in the NMSSM. 
The second solution comes again from string constructions, where 
the low-energy limit is determined by the massless string modes. Since
the massive modes are of the order of the string scale,
only trilinear couplings are present in the low-energy superpotential.
% String theory seems to be relevant for the unification of
% interactions, including gravity, and therefore this argument in favour of the
% absence of bare mass terms in the superpotential is robust.

Finally, since
the superpotential of the 
$\mu\nu$SSM contains only trilinear couplings, it
has a $Z_3$ symmetry, just like 
the NMSSM, as mentioned above.
%, under which all chiral superfields transform as 
%$\Phi\to e^{2i\pi/3}\Phi$. 
Therefore, one expects to have also a 
cosmological domain wall problem \cite{Domain walls 1,Domain walls 2} in this model. 
Nevertheless, 
the usual solution \cite{Domain walls 3} can also work in this 
case: non-renormalizable operators \cite{Domain walls 1} in the superpotential can explicitly break the dangerous $Z_3$ symmetry, lifting the degeneracy of the 
three original vacua, and this can be done without introducing hierarchy 
problems. In addition, these operators can be chosen small enough as 
not to alter the low-energy phenomenology.

% Let us also mention that terms of the type $\hat \nu^c \hat H_1 \hat H_2$ and $\hat \nu^c \hat \nu^c \hat \nu^c$ were also analysed as sources of the observed Baryon Asymmetry of the Universe (BAU) \cite{Farzan:2005ez} and of neutrino masses and tribimaximal mixing \cite{Mukhopadhyaya:2006is}, respectively. 
% In \cite{MuNu Baryogenesis}, the generation of the BAU was analysed in detail in the context of the $\mu \nu$SSM, with the interesting result that EW baryogenesis can be realized.

% The breaking of R-parity could also open the issue of fast proton decay mediated by the interchange of squarks if there are B-violating terms allowed in the superpotential. We can assume baryon triality \cite{Lepton parity and baryon triality} or that these type of operators could be prohibited by some extra gauge group coming from the compactification of a string construction while allowing the L-violating terms required by the phenomenology of this model.

% In addition, the bilinear $\mu$ term is prohibited by a discrete $Z_3$ symmetry.
%(in consequence, there is no naturalness $\mu$ problem)
% Then, one could expect, like in the NMSSM, to have a cosmological domain walls problem \cite{Domain walls 1,Domain walls 2} in this model. Nevertheless, the usual solutions to this problem \cite{Domain walls 3} will also work in this case: non-renormalizable operators \cite{Domain walls 1} in the superpotential can break explicitly the dangerous $Z_3$ symmetry, lifting the degeneracy of the three original vacua, and this can be chosen small enough as not to alter the low-energy phenomenology.

The aim of this work is to solve the above three problems adopting a different strategy.
In particular, we will add an extra $U(1)$ gauge symmetry to the gauge group of the Standard Model. 
In this way, and since all the fields of the $\mu\nu$SSM can be charged under the extra $U(1)$, all the dangerous operators could be forbidden without relying in string theory arguments, discrete symmetries or non-renormalizable operators.
Previous works using an extra $U(1)$ to solve these problems in other models, can be found in \cite{reviewlan,Peticiones}.

The outline of the paper is as follows.
% The goal of this additional symmetry is twofold. First, to forbid B-violating operators in the superpotential in order to ensure the proton stability in the context of quantum field theory without appealing to string theory arguments or discrete symmetries and second, to forbid a bilinear $\mu$ term solving the cosmological domain-walls problem. Note that in this way, the two discrete symmetries of the NMSSM (R-parity and $Z_3$) are replaced by only one additional $U(1)$ gauge symmetry that plays the same role.
In Section \ref{Section The search of models}, first we will use the extra $U(1)$ charges of the matter fields to allow the presence of the phenomenologically interesting operators, forbidding the dangerous ones. Then, we will impose the 
anomaly cancellation conditions associated to the extra $U(1)$ to constrain the values of the $U(1)$ charges.
We will see that several assignments of the $U(1)$ charges to the matter fields are viable, but in all cases the introduction of extra matter is required.
Once we have found consistent assignments (models), in Section \ref{Section Vacuum structure and compatibility with experimental constraints} we will study their phenomenology concerning the EW symmetry breaking.
We will also check that the experimental constraints on the existence of an extra gauge boson are fulfilled, as well as that correct neutrino masses can be obtained. The tree-level upper bound on the lightest Higg boson mass will also be discussed.
Finally, the conclusions are left for Section \ref{Section Conclusions}.

\section{The search of models}\label{Section The search of models}

As mentioned in the Introduction, we will work with the gauge group of the Standard Model adding an extra 
$U(1)$, 
\begin{equation}
SU(3)_C \times SU(2)_L \times U(1)_Y \times U(1)_{extra}\ . 
\end{equation}
% The matter content of the 
% $\mu\nu$SSM has then the following representations under this gauge group:
% $Q(3,2,\frac{1}{6},Q_Q)$, $u^c(\bar 3,1,-\frac{2}{3},Q_u)$, $d^c(\bar 3,1,\frac{1}{3},Q_d)$, $L(1,2,-\frac{1}{2},Q_L)$, $e^c(1,1,1,Q_e)$, $\nu^c(1,1,0,Q_{\nu^c})$, $H_1(1,2,-\frac{1}{2},Q_{H_1})$, $H_2(1,2,\frac{1}{2},Q_{H_2})$, where for simplicity we have taken the extra charges as family independent.
The matter content of the 
$\mu\nu$SSM with three families of quarks and leptons and one family of Higgses has then the following representations under this gauge group:
\begin{eqnarray}
&& Q(3,2,\frac{1}{6},Q_Q)\ , \,\,\,  u^c(\bar 3,1,-\frac{2}{3},Q_u)\ , \,\,\,
d^c(\bar 3,1,\frac{1}{3},Q_d)\ , 
\nonumber\\
&& L(1,2,-\frac{1}{2},Q_L)\ , \,\,\, e^c(1,1,1,Q_e)\ , \,\,\, \nu^c(1,1,0,Q_{\nu^c})\ ,
\nonumber\\ 
&& H_1(1,2,-\frac{1}{2},Q_{H_1})\ , \,\,\, H_2(1,2,\frac{1}{2},Q_{H_2})\ , 
\end{eqnarray}
where for simplicity we have taken the extra charges as family independent.

Now we ask the Yukawa terms, $\hat Q \hat H_1 \hat d^c, \hat Q \hat H_2 \hat u^c, \hat L \hat H_1 \hat e^c, \hat L \hat H_2 \hat \nu^c$ (that give tree-level masses to all fermions), and the effective $\mu$ term, $ \hat \nu^c \hat H_1 \hat H_2$, to be allowed in the superpotential. Since they have to be invariant 
under $U(1)_{extra}$, we can obtain five equations for the extra $U(1)$ charges. 
Using these equations we can express five charges in terms of the other three:
%, for example $Q_{H_1}, Q_{H_2}$ and $Q_{d}$:
%
\begin{eqnarray}
\label{ecuacion1}
Q_u&=& Q_{H_1}+Q_d-Q_{H_2}\ , 
%\nonumber 
\\
\label{ecuacion2}
Q_Q&=&-Q_{H_1}-Q_d\ , 
%\nonumber 
\\
\label{ecuacion3}
Q_e&=&-2Q_{H_1}\ , 
%\nonumber 
\\
\label{ecuacion4}
Q_L&=&Q_{H_1}\ , 
%\nonumber 
\\
\label{ecuacion5}
Q_{\nu^c}&=&-Q_{H_1}-Q_{H_2}\ .
%\label{ecuacion5}
%Q_{H_1}& \neq &Q_{H_2}-3Q_d \quad \mbox{to guarantee the proton stability} \nonumber \\
%Q_{H_1}& \neq &-Q_{H_2} \quad \mbox{to solve the cosmological domain wall problem} \nonumber \\
\end{eqnarray}

It is worth noticing here that equations (\ref{ecuacion2}), (\ref{ecuacion3}) and (\ref{ecuacion4}), imply that the lepton number violating terms, $\hat L \hat L \hat e^c$ and $\hat L \hat Q \hat d^c$, are automatically allowed.
Thus to avoid fast proton decay, 
the baryon number violating term, $ \hat d^c \hat d^c \hat u^c$, should be forbidden. 
Using (\ref{ecuacion1}) one obtains the following condition:
\begin{equation}
Q_{H_1} \neq Q_{H_2}-3Q_d\ .
\label{ecuacion6}
\end{equation}

Besides, to forbid the bilinear $\mu$-term, $\mu \hat H_1 \hat H_2$, one has to impose,
\begin{equation}
Q_{H_1} \neq - Q_{H_2}\ .
\label{ecuacion7}
\end{equation}
Given (\ref{ecuacion4}), this implies that the lepton number violating operator 
$\hat L \hat H_2$ is automatically forbidden.
In addition, from (\ref{ecuacion5}) one obtains that $Q_{\nu^c}\neq 0$, and, as a consequence, 
the term that generates the cosmological domain wall problem, $ \hat \nu^c \hat \nu^c \hat \nu^c$, is also automatically forbidden. 
%necessary that $Q_{H_1} \neq - Q_{H_2}$.
It is worth noticing here that a Goldstone boson does not appear from the absence of this term in the superpotential, since the $U(1)$ symmetry is gauged. As a consequence, the Goldstone boson is eaten by the $Z'$ in the process of EW symmetry breaking.
We will see in the next section that the analysis of the generalized see-saw matrix mixing neutrinos with neutralinos, in the case of three generations, is different from the usual one in the $\mu\nu$SSM because of the absence of this effective
Majorana mass term.

%  Although this effective Majorana mass term, typical of the $\mu\nu$SSM, is not present now, we will see in the next section that a generalized see-saw matrix mixing neutrinos with neutralinos can generate the correct neutrino masses.

% We can now see what happens with respect the extra $U(1)$ invariance of the rest of renormalizable terms invariant under $SU(3)_C \times SU(2)_L \times U(1)_Y$. The L-violating operators $\hat L \hat L \hat e^c$ and $\hat L \hat Q \hat d^c$ are automatically allowed by the extra $U(1)$.
% The B-violating operator, $\hat u^c \hat d^c \hat d^c$ should be forbidden. For this to happen: $Q_{H_1} \neq Q_{H_2}-3Q_d$.
% To forbid the operators $\mu \hat H_1 \hat H_2$, $\hat L \hat H_2$ and $ \hat \nu^c \hat \nu^c \hat \nu^c$ it is necessary that $Q_{H_1} \neq - Q_{H_2}$.

% We have obtained the following constraints on the extra $U(1)$ charges with the choice of terms allowed in the superpotential:
% \begin{eqnarray}\label{ecuaciones e inecuaciones}
% Q_u&=& Q_{H_1}+Q_d-Q_{H_2} \nonumber \\
% Q_Q&=&-Q_{H_1}-Q_d \nonumber \\
% Q_e&=&-2Q_{H_1} \nonumber \\
% Q_L&=&Q_{H_1} \nonumber \\
% Q_{\nu^c}&=&-Q_{H_1}-Q_{H_2}  \quad \mbox{to allow the effective $\mu$ term} \nonumber \\
% Q_{H_1}& \neq &Q_{H_2}-3Q_d \quad \mbox{to guarantee the proton stability} \nonumber \\
% Q_{H_1}& \neq &-Q_{H_2} \quad \mbox{to solve the cosmological domain wall problem} \nonumber \\
% \end{eqnarray}

Let us now impose the six anomaly cancellation conditions associated to the extra $U(1)$ gauge symmetry.
% The six conditions can be found in Table \ref{Table anomaly equations}, where a simplified notation is used. 
% \begin{table}[htb]
% $$\left |
% \begin{array}{c|c}
% \hline
% [SU(3)_C]^2 - U(1)_{extra} & \sum Y'=0 \ \mbox{(color triplet fermions only)} \\
% \hline
% [SU(2)_L]^2 - U(1)_{extra} & \sum Y'=0 \ \mbox{(doublet fermions only)} \\
% \hline
% [U(1)_Y]^2 - U(1)_{extra} & \sum Y'Y^2=0 \\
% \hline
% U(1)_Y - [U(1)_{extra}]^2 & \sum Y Y'^2=0 \\
% \hline
% [U(1)_{extra}]^3 & \sum Y'^3=0 \\
% \hline
% [Gravity]^2 - U(1)_{extra} & \sum Y'=0 \\
% \hline
% \end{array} \right|
% $$
% \caption{Anomaly cancellation conditions for the $U(1)_{extra}$ charges of the particles of the $\mu\nu$SSM.
%\cite{}. 
% The sum extends over all left-handed fermions and antifermions. $Y'$ generically denotes 
% the $U(1)_{extra}$ charges of the particles.}
% \label{Table anomaly equations}
% \end{table}

The cancellation of the anomaly $[SU(2)_L]^2 - U(1)_{extra}$ implies that the condition $\sum Q_{extra}=0$ must be fulfilled, where the sum 
extends over all left-handed fermions and antifermions, and $Q_{extra}$
generically denotes the $U(1)_{extra}$ charges of the particles (in this case only doublet fermions). Using in addition
(\ref{ecuacion2}) and (\ref{ecuacion4}),
one obtains for $Q_d$ the solution 
% \begin{equation}
% Q_d=\frac{n_H-6}{9}Q_{H_1}+\frac{n_H}{9}Q_{H_2}\ ,
% \label{qd}
% \end{equation}
$Q_d=\frac{n_H-6}{9}Q_{H_1}+\frac{n_H}{9}Q_{H_2}$,
where we have assumed in principle that the number of Higgs doublets $n_H$ is free.
Note however that if $n_H=3$ one would obtain $Q_{H_1}=Q_{H_2}-3Q_d$, which does not fulfill
(\ref{ecuacion6}) and therefore the baryon number violating operator $d^cd^cu^c$ would be allowed.
Thus we will not consider this possibility and impose $n_H\neq 3$.

The $[SU(3)_C]^2 - U(1)_{extra}$ anomaly cancellation condition, 
$\sum Q_{extra}=0$ (only for color triplet fermions), gives rise to: $3(2Q_Q+Q_u+Q_d)=0$.
But once (\ref{ecuacion1}) and (\ref{ecuacion2}) are taken into account, one obtains 
%we replace everything in terms of the three variables ($Q_{H_1},Q_{H_2},Q_d$) using (\ref{ecuaciones e inecuaciones}) we obtain 
$Q_{H_1}=-Q_{H_2}$, which does not fulfill (\ref{ecuacion7}), thus the bilinear operators would be allowed in the superpotential, spoiling the solution of the $\mu\nu$SSM to the $\mu$ problem. Then, we conclude that {\it we have to introduce exotic matter with color charge} in the spectrum to cancel this anomaly. On the other hand,
in order not to alter the anomaly cancellation conditions associated to the Standard Model gauge group,
we assume that we have $n_q$ generations of exotics which are vector-like pairs of chiral superfields with opposite $U(1)_Y$ hypercharges: $\hat q(3,1,Y_q,Q_q)$, $\hat q^c(\bar 3,1,-Y_q,Q_{q^c})$.
%in order that they do not alter the anomaly cancellation conditions among the SM gauge group. 
In addition, to avoid conflicts with experiments, the exotic quarks 
% $q$ and $q^c$
must be sufficiently heavy to not have been detected. Thus to give them masses, we add a trilinear term in the superpotential 
\begin{equation}
\lambda _q^{ijk} \hat \nu_i^c \hat q_j \hat q_k^c\ .
\end{equation}
% in order to generate an effective mass term for the color triplets when the gauge singlet sneutrinos take VEVs in the EW breaking. 
Requiring that this term is allowed by the $U(1)_{extra}$, i.e.
% \begin{equation}
% Q_{\nu^c} = -Q_{q}-Q_{q^c}\ ,
% \end{equation}
$Q_{\nu^c} = -Q_{q}-Q_{q^c}$,
and using (\ref{ecuacion5}), we obtain that relation 
\begin{equation}
Q_q+Q_{q^c}=Q_{H_1}+Q_{H_2}\ ,
\label{qextras}
\end{equation}
must be fulfilled.
Taking into account this relation together with the equation of cancellation of the $[SU(3)_C]^2 - U(1)_{extra}$ anomaly, we finally obtain that the number of families of the exotic triplets must be $n_q=3$.
% and that the relation $Q_q+Q_{q^c}=Q_{H_1}+Q_{H_2}$ has to be fulfilled (EL ORIGEN DEL 3 ESTA UN POCO CONFUSO, ADEMAS TAMBIEN SE DEBE CANCELAR LA CARGA DEL NEUTRINO CON LA DE LOS EXOTICOS Y NO SE PONE).

Let us now consider the $[Gravity]^2 - U(1)_{extra}$ anomaly cancellation condition, that is,
$\sum Q_{extra}=0$, to obtain:
%Assuming in principle that the number of Higgs doublets, $n_H$, if free, this condition is given by: 
$3(6Q_Q+3Q_u+3Q_d+2Q_L+Q_e+Q_{\nu^c})+2 n_H(Q_{H_1}+Q_{H_2})+3(3Q_q+3Q_{q^c})=0$.
Using (\ref{ecuacion1}-\ref{ecuacion5}) and (\ref{qextras}), one finally gets
% and using (\ref{ecuaciones e inecuaciones}) we obtain 
$(2n_H-3)(Q_{H_1}+Q_{H_2})=0$ which has no solution for $Q_{H_1} \neq -Q_{H_2}$ or an integer number of Higgs families. Then we conclude that we have to add more exotic matter to the spectrum in order to cancel the gravitational anomaly.
%We have seen that with only two new degrees of freedom in the extra charges ($Q_q$ and $Q_{q^c}$) we have to add more exotic matter to the spectrum. 
Since we would like to extend the model with the minimal content of matter, the simplest solution is
{\it to add extra singlets} under the Standard Model gauge group, in order not to alter the usual anomaly cancellation. 
In particular, we will add in principle $n_s$ generations of singlets $\hat s(1,1,0,Q_s)$. 
Thus the anomaly cancellation condition implies that $Q_s$ must have the value
$Q_s=\frac{3-2n_H}{n_s}(Q_{H_1}+Q_{H_2})$.
% \begin{equation}
% Q_s=\frac{3-2n_H}{n_s}(Q_{H_1}+Q_{H_2})\ .
% \label{singlets}
% \end{equation}

Now, with the $[U(1)_Y]^2 - U(1)_{extra}$ anomaly cancellation condition, 
$\sum Y^2 Q_{extra}=0$, where $Y$ generically denotes the hypercharges of the particles,
and using 
(\ref{ecuacion1}-\ref{ecuacion4}), (\ref{qextras}) and the value for $Q_d$ obtained above, one can find the 
following equation: $(9 Y_q^2+n_H-4)(Q_{H_1}+Q_{H_2})=0$. 
Since we want to forbid the bilinear terms in the superpotential, we must impose $9Y_q^2+n_H-4=0$. For 
$n_H=1, 2$ one obtains that $Y_q$ must be an irrational number. The case $n_H=3$ is excluded by the requirement of proton stability, as discussed above. For $n_H>4$ we obtain a complex value for $Y_q$.
Finally, with $n_H=4$ the $[U(1)_{extra}]^2 - U(1)_Y$ anomaly cancellation condition, $\sum Q_{extra}^2 Y =0$, implies $Q_{H_1}=-Q_{H_2}$, and the bilinear terms would be allowed in the superpotential.

We conclude with this analysis that it is not possible to cancel all the anomalies with only three new degrees of freedom ($Q_q, Q_{q^c}, Q_s$). We have checked that neither is possible with four. In particular, we have considered the following possibilities: Two types of vector-like triplets, $\hat q_1(3,1,Y_{q_1},Q_{q_1})$, $\hat q_1^c(\bar 3,1,-Y_{q_1},Q_{q^c_1})$, $\hat q_2(3,1,Y_{q_2},Q_{q_2})$ and $\hat q_2^c(\bar 3,1,-Y_{q_2},Q_{q^c_2})$; one type of vector-like triplets, $\hat q(3,1,Y_q,Q_q)$ and $\hat q^c(\bar 3,1,-Y_q,Q_{q^c})$, and two singlets with opposite hypercharge for not altering the Standard Model anomaly cancellation, $\hat s_1(1,1,Y_s,Q_{s_1})$ and $\hat s_2(1,1,-Y_s,Q_{s_2})$; one type of vector-like triplets, $\hat q(3,1,Y_q,Q_q)$ and $\hat q^c(\bar 3,1,-Y_q,Q_{q^c})$, one $SU(2)_L$ doublet with zero hypercharge, $\hat l(1,2,0,Q_l)$, and one Standard Model singlet $\hat s(1,1,0,Q_s)$; one type of vector-like triplets, $\hat q(3,1,Y_q,Q_q)$ and $\hat q^c(\bar 3,1,-Y_q,Q_{q^c})$, and one type of vector-like doublets, $\hat l(1,2,Y_l,Q_l)$ and $\hat l^c(1,2,-Y_l,Q_{l^c})$. In all these cases it is not possible to cancel all the anomalies solving all the problems discussed above, and having effective mass terms for the exotic matter.

\vspace{0.5cm}

Let us then find the simplest model that cancels all anomalies. This model adds the following exotic matter to the spectrum (with five extra charges): three generations of vector-like triplets 
\begin{equation}
\hat q(3,1,Y_q,Q_q)\ , \,\,\,\, \hat q^c(\bar 3,1,-Y_q,Q_{q^c})\ , 
\label{matter1}
\end{equation}
and $n_s$ generations of singlets 
\begin{equation}
\hat s(1,1,0,Q_s)\ , 
\label{matter3}
\end{equation}
as obtained above, and in addition $n_l$ generations of doublets
\begin{equation} 
\hat l(1,2,Y_l,Q_l)\ , \,\,\,\, \hat l^c(1,2,-Y_l,Q_{l^c})\ . 
\label{matter2}
\end{equation}
%This extra matter is vector-like under the gauge group of the SM to not alter the cancellation of the SM anomalies. 
The doublets must also be sufficiently massive as the extra triplets to have evaded the experimental detection. Then, we allow the following effective mass term in the superpotential: 
%$\hat \nu^c \hat q \hat q^c$ and 
\begin{equation}
\lambda_l^{ijk}\hat \nu^c_i \hat l_j \hat l^c_k\ . 
\end{equation}
As a consequence, the following condition is obtained: 
%and we obtain two equations for the extra charges: $Q_q+Q_{q^c}=Q_{H_1}+Q_{H_2}$ and 
\begin{equation}
Q_l+Q_{l^c}=Q_{H_1}+Q_{H_2}\ .
\label{eles}
\end{equation}
% Let us study the anomaly cancellation equations with the help of (\ref{ecuaciones e inecuaciones}):\\
% \begin{itemize}
%  \item $[SU(3)_C]^2 - U(1)_{extra}$ gives $n_q=3$.\\
% \end{itemize}

In addition, the $[SU(2)_L]^2 - U(1)_{extra}$ anomaly cancellation gives rise to:
\begin{equation}
Q_d=\frac{n_H+n_l-6}{9}Q_{H_1} + \frac{n_H+n_l}{9}Q_{H_2}\ . 
\end{equation}
Note that the requirement (\ref{ecuacion6})
%$Q_{H_1} \neq    Q_{H_2}-3Q_d$ to forbid the B-violating operator means 
implies that $n_H+n_l \neq 3$.\\

From the $[Gravity]^2 - U(1)_{extra}$ anomaly cancellation we obtain the value of the extra charge of the singlet: 
\begin{equation}
Q_s=\frac{3-2n_H-2n_l}{n_s}(Q_{H_1}+Q_{H_2})\ .
\end{equation}

After replacing all the variables in the $[U(1)_Y]^2 - U(1)_{extra}$ anomaly cancellation condition we are left with the equation 
\begin{equation}
18\ Y_q^2+4\ n_l\ Y_l^2=8-n_l-2\ n_H\ . 
\end{equation}
The left side of the equation is a sum of positive quantities so we obtain an upper bound on the number of generations  $n_l+2n_H \leq 8$. We can study this equation searching for reasonable rational values of the hypercharges. The results are given in Table \ref{Table Hypercharges}.

%\begin{table}
%$$\left |
%\begin{array}{c|c|c|c}
%\hline
%n_H & n_l & Y_q & Y_l \\
%\hline
%1 & 3 & \pm \frac{4}{10} & \pm \frac{1}{10} \\
%\hline
%1 & 3 & 0 & \pm \frac{1}{2} \\
%\hline
%1 & 4 & \pm \frac{1}{3} & 0 \\
%\hline
%1 & 4 & \pm \frac{1}{9} & \pm \frac{1}{3} \\
%\hline
%2 & 2 & \pm \frac{1}{5} & \pm \frac{2}{5} \\
%\hline
%2 & 2 & \pm \frac{1}{3} & 0 \\
%\hline
%3 & 1 & \pm \frac{2}{9} & \pm \frac{1}{6} \\
%\hline
%2 & 2 & 0 & \pm \frac{1}{2} \\
%\hline
%3 & 1 & 0 & \pm \frac{1}{2} \\
%\hline
%\end{array} \right|
%$$
\begin{table}
$$\left |
\begin{array}{c|c|c|c|c|c|c|c|c|c}
\hline
n_H & 1 & 1 & 1 & 1 & 2 & 2 & 3 & 2 & 3 \\
\hline
n_l & 3 & 3 & 4 & 4 & 2 & 2 & 1 & 2 & 1 \\
\hline
Y_q & \pm \frac{2}{5} & 0 & \pm \frac{1}{3} & \pm \frac{1}{9} & \pm \frac{1}{5} & \pm \frac{1}{3}& \pm \frac{2}{9} & 0 & 0 \\
\hline
Y_l & \pm \frac{1}{10} & \pm \frac{1}{2} & 0 & \pm \frac{1}{3} & \pm \frac{2}{5} & 0 & \pm \frac{1}{6} & \pm \frac{1}{2} & \pm \frac{1}{2} \\
\hline
\end{array} \right |
$$
\caption{Number of generations of Higgses and extra doublets, and hypercharges that 
solve the $[U(1)_Y]^2 - U(1)_{extra}$ anomaly equation.}
\label{Table Hypercharges}
\end{table}
%\end{itemize}

The equation associated to $[U(1)_{extra}]^2 - U(1)_Y$ is quadratic in the extra charges:
\begin{eqnarray}
&& 3(6 Y_Q Q_Q^2+3Y_u Q_u^2+3Y_d Q_d^2+2Y_LQ_L^2+Y_eQ_e^2+Y_\nu Q_{\nu^c}^2)+n_H(2Y_{H_1} Q_{H_1}^2+2Y_{H_2} Q_{H_2}^2)
\nonumber\\
&& + n_q(3Y_q Q_q^2-3Y_q Q_{q^c}^2)+n_l(2Y_l Q_l^2-2Y_l Q_{l^c}^2)+n_sY_s Q_s^2=0\ .
\end{eqnarray} 
After substituting all the extra charges known, we obtain the value of $Q_q$ in terms of $Q_{H_1}, Q_{H_2}$, 
and $Q_l$. Using (\ref{qextras}) 
%$Q_q+ Q_{q^c}=Q_{H_1}+Q_{H_2}$ 
we also obtain $Q_{q^c}$ in terms of $Q_{H_1}, Q_{H_2}$, and $Q_l$.

Finally, the equation associated to $[U(1)_{extra}]^3$ 
is cubic in the extra charges, $\sum Q_{extra}^3=0$. 
We study this equation for each set of $(n_H,n_l,Y_q,Y_l)$ of Table \ref{Table Hypercharges}. We obtain $Q_l$ in terms of $Q_{H_1}$ and $Q_{H_2}$. Using (\ref{eles}), we also obtain $Q_{l^c}$ in terms of $Q_{H_1}$ 
and $Q_{H_2}$. The only cases that give rise to rational values for $Q_l$ and $Q_{l^c}$ have the following number of generations:
\begin{equation}
n_H=1\ , \,\,\, n_l=3\ , \,\,\,  n_s=6\ ,
\end{equation}
with
\begin{equation}
Y_q=\pm \frac{2}{5}\ , \,\,\, Y_l=\frac{1}{10}\ ,
\end{equation}
and two distinct solutions for $Q_l$ as shown in the right side of 
Table \ref{Extra charges exotics}, or 
\begin{equation}
Y_q=0\ , \,\,\, Y_l=\frac{1}{2}\ ,
\end{equation}
and two distinct solutions for $Q_q$, as shown also in the right side of Table \ref{Extra charges exotics}.

It is worth noticing here that, although at the end we are left with the six different solutions (models) discussed above, we will see in the next section that all of them give rise to the same phenomenology at low energies. This is because the six models only differ in the extra charges and hypercharges of the exotic matter, and this matter does not play any role in the EW breaking.

We have then obtained all the extra charges in terms of two of them, $Q_{H_1}$ and $Q_{H_2}$. For rational values of $Q_{H_1}$ and $Q_{H_2}$ we obtain rational values for the rest of extra charges. For definiteness, we add two additional conditions for the complete determination of the extra charges. First, we impose that the bases of the hypercharge $Y$ and the extra charge $Q$ are orthogonal, i.e. $Tr[YQ]=0$. 
%This can be justified supposing that the extra charge and the hypercharge are both generators of a simple underlying gauge group and one expects them to be 
%orthogonal \cite{Bases ortogonales}. 
This implies $Q_{H_2}=6 \ Q_{H_1}$. 
Second, following \cite{Japoneses} we impose the normalization condition for the extra charges $Tr[Y^2]=Tr[Q^2]$. This condition is in fact non physical since the relevant quantity is the product of the extra gauge coupling constant $g'_1$ by the normalization factor. 
%The normalization factor is irrational but this is irrelevant. The only thing that is not allowed is to have some extra charges rational and others irrational if we want the possibility of embedding the model into a grand unified framework with a simple gauge group. 
From this condition we obtain the value of $Q_{H_1}$ and, consequently, the values of all the extra charges. We show these values in Tables \ref{Extra charges exotics} and \ref{Extra charges}, where
the normalization factor is given by $N=\sqrt{\frac{3}{2426}}$.
% We also show in Table \ref{Extra charges exotics} the extra charges and hypercharges of the exotic matter for the six models found here.
% \begin{table}[htb]
% $$\left |
% \begin{array}{c|c|c|c|c}
% \hline
% Q_{H_1}=3 \ R & Q_{H_2}=18 \ R & Q_Q=-\frac{31}{3} \ R & Q_u=- \frac{23}{3} R & Q_d=\frac{22}{3} \ R \\ 
% \hline
% Q_L=3 \ R & Q_e=-6 \ R & Q_{\nu^c} = -21 \ R & Q_s=- \frac{35}{2} \ R \\
% \hline
% \end{array} \right|
% $$
% \caption{Numerical values of the extra charges of the particle content of the $\mu \nu$SSM and the extra singlets.}
% \label{Extra charges}
% \end{table}

\begin{table}[htb]
{\normalsize
$$\left |
\begin{array}{c|c|c|c|c}
\hline
Q_q &  Q_{q^c} & Q_l &  Q_{l^c} \\
\hline
\frac{257}{30}N& \frac{373}{30}N & \frac{19}{15}N & \frac{296}{15}N & \mbox{Model 1:} \ Y_q=\frac{2}{5}, \ Y_l=\frac{1}{10}, \ Q_l=\frac{1}{45}(-5Q_{H_1}+4Q_{H_2}) \\
\hline
\frac{173}{30}N & \frac{457}{30}N & \frac{271}{15} N & \frac{44}{15}N & \mbox{Model 2:} \ Y_q=\frac{2}{5}, \ Y_l=\frac{1}{10}, \ Q_l=\frac{1}{45}(31Q_{H_1}+40Q_{H_2}) \\
\hline
\frac{373}{30}N & \frac{257}{30} N & \frac{19}{15}N & \frac{296}{15}N & \mbox{Model 3:} \ Y_q=-\frac{2}{5}, \ Y_l=\frac{1}{10}, \ Q_l=\frac{1}{45}(-5Q_{H_1}+4Q_{H_2}) \\
\hline
\frac{457}{30}N & \frac{173}{30}N & \frac{271}{15} N & \frac{44}{15}N & \mbox{Model 4:} \ Y_q=-\frac{2}{5}, \ Y_l=\frac{1}{10}, \ Q_l=\frac{1}{45}(31Q_{H_1}+40Q_{H_2}) \\
\hline
\frac{7}{2} N & \frac{35}{2} N & \frac{19}{3} N & \frac{44}{3} N & \mbox{Model 5:} \ Y_q=0, \ Y_l=\frac{1}{2}, \ Q_q=\frac{1}{6}(Q_{H_1}+Q_{H_2}) \\
\hline
\frac{35}{2}N & \frac{7}{2} N & \frac{19}{3} N & \frac{44}{3} N & \mbox{Model 6:} \ Y_q=0, \ Y_l=\frac{1}{2}, \ Q_q=\frac{5}{6}(Q_{H_1}+Q_{H_2}) \\
\hline
\end{array} \right |
$$
}
\caption{Values of the $U(1)_{extra}$ charges of the extra triplets and doublets added to the Standard Model spectrum of the $\mu\nu$SSM, for the six solutions of the $[U(1)_{extra}]^3$ anomaly cancellation condition.}
\label{Extra charges exotics}
\end{table}

\begin{table}[htb]
$$\left |
\begin{array}{c|c|c|c|c}
\hline
Q_{H_1}=3 \ N & Q_{H_2}=18 \ N & Q_Q=-\frac{31}{3} \ N & Q_u=- \frac{23}{3} N & Q_d=\frac{22}{3} \ N \\ 
\hline
Q_L=3 \ N & Q_e=-6 \ N & Q_{\nu^c} = -21 \ N & Q_s=- \frac{35}{2} \ N \\
\hline
\end{array} \right|
$$
\caption{Values of the $U(1)_{extra}$ charges for the Standard Model content of the $\mu \nu$SSM and for the extra singlets.}
\label{Extra charges}
\end{table}

Summarizing, we have found six interesting models with the following exotic matter: three generations of vector-like color triplets with respect to the Standard Model gauge group (\ref{matter1}), 
%$\hat q(3,1,Y_q,Q_q)$ and $\hat q^c(\bar 3,1,-Y_q,Q_{q^c})$, 
three generations of $SU(2)_L$ doublets (\ref{matter2}),
%$\hat l(1,2,Y_l,Q_l)$ and $\hat l^c(1,2,-Y_l,Q_{l^c})$, 
and six\footnote{If preferred, one could imagine that those six generations are in fact $3+3$ generations that could be distinguished by some high-energy extra $U(1)$ gauge group, perhaps coming from the compactification of a string model \cite{old}. } Standard Model singlets (\ref{matter3}).
%$\hat s(1,1,0,Q_s)$. 
The superpotential is given by:
\begin{eqnarray}\label{superpotencial munu u1extra} W & =& 
\epsilon _{ab}(Y_u^{ij} \hat H_2^b \hat Q_i^a \hat u_j^c+Y_d^{ij} \hat H_1^a \hat Q_i^b \hat d_j^c+Y_e^{ij} \hat H_1^a \hat L_i^b \hat e_j^c + Y_\nu ^{ij} \hat H_2^b \hat L_i^a \hat \nu _j^c)  -  \epsilon _{ab} \lambda^i \hat \nu _i^c \hat H_1^a \hat H_2^b \nonumber \\ 
& +& 
\epsilon _{ab}( \lambda '^{ijk} \hat Q_i^a \hat L_j^b \hat d_k^c + \lambda^{'''ijk} \hat L_i^a \hat L_j^b \hat e_k^c) + \lambda _q^{ijk} \hat \nu _i^c \hat q_j \hat q_k^c + \epsilon _{ab} \lambda _l^{ijk} \hat \nu _i^c \hat l_j^a ( \hat l_k^b)^c \ .
\end{eqnarray}
The extra singlets do not have couplings in the superpotential. The extra triplets and doublets have effective mass terms in the superpotential avoiding conflicts with the experimental searches of exotic matter. The singlets, as they do not couple in the superpotential, and only interact through their extra $U(1)$ charge, do not need to be massive for escaping detection since the value of the lower bound on the mass of an extra $Z$ is quite large.

Let us now make a comment on the hypercharges of the extra matter. In these models, the hypercharges of the exotic matter lead to non-standard fractional electric charges. This issue has been discussed for example in \cite{Fractional electric charge Carlos}, and references therein. In the case of extra triplets, they could form color-neutral fractionally charged states since the triplets can bind. 
%In principle, the existence of stable charged states could create conflicts with cosmological bounds. 
%Thermal production of these particles would overclose the Universe unless their masses are below a few TeV \cite{Fractional 1,Fractional 2}. 
%In models with non-standard extra triplets, 
The 
%lightest color-neutral fractionally charged state
lightest of these states will be stable due to electric charge conservation.
As pointed out in \cite{dine}, the estimation of its relic abundance contradicts limits on the existence of fractional charge in matter which is less than $10^{-20}$ per nucleon \cite{Fractional 2}. Thus, avoiding such fractionally charged states is necessary. A possible mechanism to carry it out is inflation. Inflation would dilute these particles. The reheating temperature $T_{RH}$ should be low enough not to produce them again. This reheating temperature must be smaller than $10^{-3}$ times the mass of the particle \cite{T reheating}, so in our case $T_{RH}<1$ GeV. This, in principle is possible since the only constraint on this temperature is to be larger than $1$ MeV not to spoil the successful nucleosynthesis predictions.

%This model is the simplest model found which prohibits B-violating terms in the superpotential and solves the cosmological domain walls problem. 
Finally, let us recall that the models where $R$-parity is conserved still need some fine-tuning to agree with the experimental bounds on the proton lifetime. This is because $R$-parity does not forbid non-renormalizable dimension five operators that break baryon or lepton number, and could produce too fast proton decay if the couplings are of order one \cite{Dimension 5 operators}. 
We have checked that although in the model analyzed here, there are 43 non-renormalizable dimension five baryon number violating operators allowed by the gauge symmetry of the Standard Model, such as for example $\hat Q \hat Q \hat Q \hat L$, $\hat u^c \hat u^c \hat d^c \hat e^c$ or $\hat Q \hat Q \hat Q \hat H_1$, all of them turn out to be forbidden by the extra $U(1)$. 
% In this sense, the extra $U(1)$ gauge symmetry is more successful than R-parity in avoiding the problem of proton decay. It is then clear that the $\mu\nu$SSM with an extra $U(1)$ is safe from constraints from the non-observation of proton decay at Super-Kamiokande \cite{Kobayashi proton decay}. 

%The number of generations of the exotic matter introduced is three, just like the number of generations of the SM except the case of the singlets, which are six times replicated. Anyway, we can think that the six generations are in fact, $3+3$ generations and can be distinguished by some other extra $U(1)$ gauge group coming from the compactification of a string model. 

\section{Electroweak breaking and experimental constraints}\label{Section Vacuum structure and compatibility with experimental constraints}

% The superpotential of the model is given by the following expression:
% \begin{eqnarray}\label{superpotencial munu u1extra} W & =& 
% \epsilon _{ab}(Y_u^{ij} \hat H_2^b \hat Q_i^a \hat u_j^c+Y_d^{ij} \hat H_1^a \hat Q_i^b \hat d_j^c+Y_e^{ij} \hat H_1^a \hat L_i^b \hat e_j^c + Y_\nu ^{ij} \hat H_2^b \hat L_i^a \hat \nu _j^c)  -  \epsilon _{ab} \lambda^i \hat \nu _i^c \hat H_1^a \hat H_2^b \nonumber \\ 
% & +& 
% \epsilon _{ab}( \lambda '^{ijk} \hat Q_i^a \hat L_j^b \hat d_k^c + \lambda ^{ijk} \hat L_i^a \hat L_j^b \hat e_k^c) + \lambda _q^{ijk} \hat \nu _i^c \hat q_j \hat q_k^c + \epsilon _{ab} \lambda _l^{ijk} \hat \nu _i^c \hat l_j^a ( \hat l_k^b)^c \
% \end{eqnarray}
% where $i,j,k=1,2,3$ are family indices and $a,b$ are $SU(2)$ indices.
The gauge symmetry discussed in the previous section $SU(3)_C \times SU(2)_L \times U(1)_Y \times U(1)_{extra}$ has to be spontaneously broken to $SU(3)_C \times U(1)_{e.m.}$. To discuss this breaking we have first to calculate the neutral scalar potential, which is the sum of three contributions: F-terms, D-terms and soft terms.
%All the terms in the superpotential (\ref{superpotencial munu u1extra}) are trilinear terms and the couplings are adimensional. As a consequence, there are not naturalness problems. 
Working in the framework of gravity-mediated supersymmetry breaking, and taking into account the superpotential (\ref{superpotencial munu u1extra}), the latter are given by: 
\begin{eqnarray} \label{L soft munu u1extra}
\mathcal L _{soft} & =&
\frac{1}{2}( M_3 \tilde \lambda _3 \tilde \lambda _3+ M_2 \tilde \lambda _2 \tilde \lambda _2 +  M_1 \tilde \lambda _1 \tilde \lambda _1 +  M'_1 \tilde \lambda '_1 \tilde \lambda'_1+h.c.) \nonumber \\ 
& -&
\epsilon _{ab}[(A_u Y_u)^{ij}H_2^b \tilde Q^a_i \tilde u_j^c+(A_d Y_d)^{ij} H_1^a \tilde Q_i^b \tilde d_j^c+(A_e Y_e)^{ij}H_1^a \tilde L_i^b \tilde e_j^c+(A_\nu Y_\nu)^{ij}H_2^b \tilde L_i^a \tilde \nu _j^c \nonumber \\ 
& +&
(A_{\lambda'} \lambda')^{ijk}\tilde Q_i^a \tilde L_j^b \tilde d_k^c+(A_{\lambda'''} \lambda''')^{ijk} \tilde L_i^a \tilde L_j^b \tilde e_k^c-(A_\lambda \lambda )^i \tilde \nu _i^c H_1^a H_2^b + (A_{\lambda_l} \lambda _l)^{ijk} \tilde \nu _i^c \tilde l_j^a \tilde l_b^{kc} +h.c.] \nonumber \\ 
& -& 
[(A_{\lambda_q} \lambda _q)^{ijk} \tilde \nu _i^c \tilde q_j \tilde q_k^c+h.c] -[(M_{\tilde Q}^2)^{ij} \tilde Q_i^{a*} \tilde Q_j^a+(M_{\tilde u^c}^2)^{ij}\tilde u_i^{c*} \tilde u_j^c+(M_{\tilde d^c}^2)^{ij} \tilde d_i^{c*} \tilde d_j^c \nonumber \\
& +&
(M_{\tilde L}^2)^{ij} \tilde L_i^{a*} \tilde L_j^a+(M_{\tilde e^c}^2)^{ij} \tilde e_i^{c*} \tilde e_j^c+M_{H_1}^2 H_1^{a*}H_1^a+M_{H_2}^2 H_2^{a*}H_2^a+(M_{\tilde \nu ^c}^2)^{ij} \tilde \nu _i^{c*} \tilde \nu _j^c \nonumber \\ 
& +&
(M_{\tilde s}^2)^{ij} \tilde s_i^* \tilde s_j+(M_{\tilde q}^2)^{ij} \tilde q_i^* \tilde q_j+(M_{\tilde q^c}^2)^{ij} \tilde q_i^{c*} \tilde q_j^c+(M_{\tilde l}^2)^{ij} \tilde l_i^* \tilde l_j+ (M_{\tilde l^c}^2)^{ij} \tilde l_i^{ac*} \tilde l_j^{ac}] \ .
\end{eqnarray}
% We would like to study the breaking of the gauge symmetry: $$SU(3)_C \times SU(2)_L \times U(1)_Y \times U(1)_{extra} \rightarrow SU(3)_C \times U(1)_{e.m.}$$
% For this, we have first to calculate the neutral scalar potential, which is the sum of three contributions: F-terms, D-terms and soft terms. 
Once the EW symmetry is spontaneously broken, the neutral scalars develop in general the following VEVs: 
\begin{equation}
\langle H_1^0 \rangle=v_1, \;\;\;\;\langle H_2^0 \rangle =v_2,\;\;\;\; \langle \tilde \nu_i \rangle=\nu_i\ \;\;\;\;
\langle \tilde \nu_i ^c \rangle=\nu_i ^c\ . 
\end{equation}
We have checked that the neutral components of the exotic matter do not take VEVs in a wide region of the parameter space, where we will concentrate. In what follows, it will be enough for our purposes to neglect mixing between generations 
in (\ref{superpotencial munu u1extra}) and (\ref{L soft munu u1extra}), and to assume that only one generation of sneutrinos gets VEVs, $\nu$, $\nu^c$. The extension of the analysis to all generations is straightforward, and the conclusions are similar. The expression of the neutral scalar potential is then given by:
\begin{eqnarray} \label{VEV potencial extra u1}
<V^0> &= &
\frac{1}{8}(g_1^2+g_2^2)(|v_1|^2+|\nu|^2-|v_2|^2)^2 \nonumber \\ 
&+ &
\frac{1}{2}  g^{\prime2}_1(Q_{H_1}|v_1|^2+Q_{H_2}|v_2|^2+Q_L|\nu|^2+Q_{\nu ^c}|\nu ^c|^2)^2 \nonumber \\ 
&+ &
|Y_\nu|^2(|v_2|^2|\nu ^c|^2+|v_2|^2|\nu|^2+|\nu |^2|\nu ^c|^2) \nonumber \\
&+ &
|\lambda |^2(|v_1|^2|v_2|^2+|\nu ^c|^2|v_2|^2+|\nu ^c|^2|v_1|^2) \nonumber \\
&+ &
(-\lambda  Y_\nu ^* v_1 \nu ^* |v_2|^2-\lambda  Y_\nu ^* v_1 \nu ^* |\nu ^c|^2 +h.c.) \nonumber \\ 
&+ &
M_{\tilde L} ^2 |\nu|^2+M_{\tilde \nu ^c}^2|\nu ^c|^2+M_{H_1}^2|v_1|^2+M_{H_2}^2|v_2|^2 \nonumber \\
&+ &
(A_\nu Y_\nu v_2 \nu \nu ^c-A_\lambda \lambda  \nu ^c v_1 v_2+h.c.)\ .
\end{eqnarray}
We also assume, for simplicity, that there is not CP violation in the scalar sector and we take all the parameters and VEVs real in (\ref{VEV potencial extra u1}). The four minimization conditions with respect to the 
VEVs $v_1, \ v_2, \ \nu^c, \ \nu $, are:
\begin{eqnarray*}
&& \frac{1}{4}(g_1^2+g_2^2)(v_1^2+\nu ^2 -v_2^2)v_1+g_1^{\prime2}(Q_{H_1}v_1^2+Q_{H_2}v_2^2+Q_L \nu ^2+Q_{\nu ^c} {\nu ^c}^2)Q_{H_1} v_1 \nonumber \\
& +& \lambda ^2 v_1(v_2^2+{\nu ^c}^2)+M_{H_1}^2v_1-\lambda  Y_\nu \nu |v_2|^2-\lambda  Y_\nu \nu |\nu ^c|^2-A_\lambda \lambda  \nu ^c v_2=0\ , \nonumber 
\end{eqnarray*}
\begin{eqnarray*}
& -& \frac{1}{4}(g_1^2+g_2^2)(v_1^2+\nu ^2-v_2^2)v_2+g_1^{\prime2}(Q_{H_1}v_1^2+Q_{H_2}v_2^2+Q_L \nu ^2+Q_{\nu ^c} {\nu ^c}^2)Q_{H_2}v_2 \nonumber \\
& +& Y_\nu ^2 v_2(\nu ^2+{\nu ^c}^2)+\lambda ^2 v_2(v_1^2+{\nu ^c}^2)+M_{H_2}^2v_2-2\lambda  Y_\nu v_1 \nu v_2 \nonumber \\
& +& A_\nu Y_\nu \nu \nu ^c-A_\lambda \lambda  \nu ^c v_1=0\ , \nonumber 
\end{eqnarray*}
\begin{eqnarray*}
&& g_1^{\prime2}(Q_{H_1}v_1^2+Q_{H_2}v_2^2+Q_L \nu ^2+Q_{\nu ^c} {\nu ^c}^2)Q_{\nu ^c}\nu ^c+Y_\nu ^2 \nu ^c (v_2^2+\nu ^2) -A_\lambda \lambda  v_1 v_2\nonumber \\
& +& \lambda ^2 \nu ^c (v_1^2+v_2^2)+M_{\tilde \nu ^c}^2 \nu ^c-2\lambda  Y_\nu v_1 \nu \nu ^c+A_\nu Y_\nu v_2 \nu =0\ , \nonumber
\end{eqnarray*}
\begin{eqnarray}
&& \frac{1}{4}(g_1^2+g_2^2)(v_1^2+\nu ^2-v_2^2)\nu +g_1^{\prime2}(Q_{H_1}v_1^2+Q_{H_2}v_2^2+Q_L\nu ^2+Q_{\nu ^c}{\nu ^c}^2)Q_L \nu \nonumber \\
& +& Y_\nu ^2 \nu (v_2^2+{\nu ^c}^2)+M_{\tilde L} ^2 \nu -\lambda  Y_\nu v_1 v_2^2-\lambda  Y_\nu v_1 {\nu ^c}^2+A_\nu Y_\nu v_2 \nu ^c=0\ . 
\label{ecuaciones minimo}
\end{eqnarray}
%In the context of R-parity breaking models with extra singlets \cite{Masiero y Valle}, the VEV of the left-handed sneutrino, $\nu$, is in general small. 
Notice that in the last equation in (\ref{ecuaciones minimo}) $\nu \rightarrow 0$ as $Y_\nu \rightarrow 0$, and since the coupling $Y_\nu$ determines the Dirac mass for the neutrinos, $m_D \equiv Y_\nu v_2$, then
$Y_{\nu}\sim 10^{-6}$, and therefore
$\nu$ has to be very small. The smallness of the left-handed sneutrino VEVs for a correct description of the neutrino sector in the $\mu \nu$SSM, compatible with current data, has been proved in \cite{MuNu Indian,MuNu Hirsch,MuNu SCPV}. 

%Using this rough argument, 
We can now approximate the minimization equations neglecting the values of $\nu$ and $Y_\nu$, and we are left with only three equations. 
%Note that we are not assuming any hypothesis on the soft masses at high energy and we take the soft parameters as free parameters at the EW scale. We can s
Solving the minimization conditions for the soft masses in terms of the extra charges, coupling constants, VEVs, and the parameters $\lambda $ and $A_\lambda \lambda $, one obtains:
%We will impose the experimental constraints on the existence of a new gauge boson. They define a region in the space of the VEVs and we will translate these constraints on the soft masses space using the minimization conditions.
%The three minimization conditions read:
\begin{eqnarray} \label{3 ecs minim munu u1 msoft despejadas}
M_{H_1}^2 &=& -\frac{1}{4}(g_1^2+g_2^2)(v_1^2-v_2^2)-g_1^{\prime2}(Q_{H_1}v_1^2+Q_{H_2}v_2^2+Q_{\nu ^c} {\nu ^c}^2)Q_{H_1} \nonumber \\ 
&-& \lambda ^2(v_2^2+{\nu ^c}^2)+A_\lambda \lambda  \nu ^c \frac{v2}{v1}\ , \nonumber \\
M_{H_2}^2 &=& \frac{1}{4}(g_1^2+g_2^2)(v_1^2-v_2^2)-g_1^{\prime2}(Q_{H_1}v_1^2 +Q_{H_2}v_2^2 +Q_{\nu ^c} {\nu ^c}^2)Q_{H_2} \nonumber \\
&-& \lambda  ^2 (v_1^2+{\nu ^c}^2)+A_\lambda \lambda  \nu ^c \frac{v1}{v2}\ , \nonumber \\
M_{\tilde \nu ^c}^2 &=& -g_1^{\prime2}(Q_{H_1}v_1^2+Q_{H_2}v_2^2+Q_{\nu ^c} {\nu ^c}^2)Q_{\nu ^c} -\lambda ^2(v_1^2+v_2^2)+A_\lambda \lambda  \frac{v_1 v_2}{\nu ^c}\ .
\end{eqnarray}
Note that these equations are equivalent (substituting $\nu^c$ by the VEV of a singlet scalar) to the minimization conditions for the $U(1)$SSM models \cite{extras1,Japoneses}, where correct EW breaking is known to take place.

On the other hand, the VEVs have to satisfy several phenomenological constraints. First, the mass of the $W$ boson, 
$M_{W}=\frac{1}{2}g_2^2 (v_1^2+v_2^2+\nu ^2)$, is well determined, leading to $(v_1^2+v_2^2) \simeq (174 \ \text{GeV})^2$ when $\nu$ is neglected. Second, the $Z$ boson of the Standard Model and the $Z'$ boson associated to the $U(1)_{extra}$ are mixed with a mass-squared matrix given by:
\begin{eqnarray}
\left(
\begin{array}{cc}
M_Z^2 & M_{ZZ'}^2 \\
M_{ZZ'}^2 & M_{Z'}^2\ 
\end{array}
\right)\ ,
\end{eqnarray}
where the entries are functions of the VEVs, gauge coupling constants and extra charges,
 \begin{eqnarray}
M_Z^2 &=& \frac{1}{2}(g_1^2+g_2^2)(v_1^2+v_2^2)\ , \nonumber \\
M_{Z'}^2 &=& 2g_1^{\prime2}(Q_{H_1}^2v_1^2+Q_{H_2}^2v_2^2+Q_{\nu ^c}^2 {\nu ^c}^2)\ , \nonumber \\
M_{ZZ'}^2 &=& g'_1\sqrt{g_1^2+g_2^2}(-Q_{H_1}v_1^2+Q_{H_2}v_2^2)\ .
\end{eqnarray}
Diagonalizing this matrix one obtains the mass eigenstates. The experimental constraints imply the following bound \cite{Cota mezcla} for
the mixing parameter
\begin{equation}
R=\dfrac{(M^2_{ZZ'})^2}{M_Z^2M^2_{Z'}}\leq 10^{-3}\ .
\end{equation}
In addition, the mass of the heaviest eigenstate should be larger than about $1 \ \text{TeV}$ \cite{Cota masa Z'}.
%and the mixing parameter $R=\dfrac{(M^2_{ZZ'})^2}{M_Z^2M^2_{Z'}}$ should be smaller than $10^{-3}$ \cite{Cota mezcla}. 
If we also ask the heaviest eigenstate to be lighter than $2000 \ \text{GeV}$ in order not to have a very large fine-tuning (and for the $Z'$ to be discovered at present accelerator experiments), then
%We put an upper bound on $tan \ \beta$ as well.
%The inequalities that 
%we obtain the constraint
%define an allowed region in the parameter space of the VEVs:
\begin{eqnarray}
% && R=\frac{(Q_{H_2} v_2^2-Q_{H_1}v_1^2)^2}{(v_1^2+v_2^2)(Q_{H_1}^2v_1^2+Q_{H_2}^2v_2^2+Q_{\nu ^c}^2{\nu ^c}^2)} \leq 10^{-3} \nonumber \\
&& (1000)^2 \leq \frac{1}{4}(g_1^2+g_2^2)(v_1^2+v_2^2)+g_1^{\prime2}(Q_{H_1}^2v_1^2+Q_{H_2}^2v_2^2+Q_{\nu ^c}^2 {\nu^c}^2) \nonumber \\
&& +[\frac{1}{16}(g_1^2+g_2^2)^2(v_1^2+v_2^2)^2+g_1^{\prime4}(Q_{H_1}^2v_1^2+Q_{H_2}^2v_2^2+Q_{\nu^c}^2 {\nu^c}^2)^2 \nonumber \\
&& -\frac{1}{2}(g_1^2+g_2^2)g_1^{\prime2}(v_1^2+v_2^2)(Q_{H_1}^2v_1^2+Q_{H_2}^2v_2^2+Q_{\nu^c}^2{\nu^c}^2) \nonumber \\
&& +g_1^{\prime2}(g_1^2+g_2^2)(-Q_{H_1}v_1^2+Q_{H_2}v_2^2)^2]^{1/2} \leq (2000)^2\ .
\end{eqnarray}

From the above equations it is obvious that the six models found in the previous section give rise to the same phenomenology at low energies, since 
they only differ in the extra charges and hypercharges of the exotic matter, and this matter does not play any role in the EW breaking.
% Note that the phenomenology analyzed here does not depend on the six models found since they only differ from one to another on the hypercharges and extra charges of the exotic matter and it does not play any role on the electroweak breaking neither in the $Z'$ constraints.

In order to study the solutions of the equations,
%For the rest of the soft parameters that are involved in the minimization equations (\ref{3 ecs minim munu u1 msoft despejadas}) 
we assume the following reasonable values for the parameters: $A_\lambda \lambda = 0.1 \ \text{TeV}$ and $ \lambda  = 0.1, 0.3$. 
%For definiteness, 
For the sake of definiteness we also take 
%take the value of the extra gauge coupling constant equal to the gauge coupling constant of the $U(1)_Y$: 
$g'_1=g_1$,
%[together with the extra charges normalization condition $Tr[Y'^2]=Tr[Y^2]$ like in \cite{Japoneses}], 
$1 < \tan\beta=\frac{v_2}{v_1} <  35$, and we work in the parameter space $(\nu^c, \tan \beta)$.
Once imposed 
the experimental constraints on the existence of a new gauge boson $Z'$, we have checked that 
the effect of the bound on the $Z-Z'$ mixing is more important than the bounds on the mass of the heaviest eigenstate,
although it is still possible to find wide allowed regions.
The former experimental constraint implies a lower bound on the VEV of the right-handed sneutrino $\nu^c$, depending on the value of $\tan \beta$. In particular,
for $\lambda  = 0.3$ and $\tan\beta=1$, $\nu^c$ must be larger than 2 TeV.
For increasing values of $\tan \beta$, the lower bound on $\nu^c$ increases since it is more difficult to suppress the $Z-Z'$mixing. For example, for $\tan\beta=3$ (7), one obtains that $\nu^c$ must be larger than about 4 (4.6) TeV.
For $\tan \beta$ larger than 7, the lower bound on $\nu^c$ practically does not vary.
Similar results are obtained for $\lambda  = 0.1$, although in this case a tachyonic region appears and we always need values of $\nu^c$ larger than 2.5 TeV.

% , and imposing 
% the experimental constraints on the existence of a new gauge boson $Z'$, 
% it is still possible to find wide allowed regions.
%in Figures \ref{VEVs01},\ref{VEVs03} (light/blue region)
% for the cases $ \lambda =0.1$ and $ \lambda  =0.3$ respectively. We have verified that the critical points of the potential are minima analyzing the Hessian constructed with the second derivatives of the potential. In the case $\lambda=0.1$ a large tachyonic region appears.

%\begin{figure}[htb]
% \begin{center}
%\hspace*{1cm}
%     \epsfig{file=figures/VEVs01.ps ,height=10.0cm,angle=0}
%\caption{Allowed region by the $Z'$ experimental constraints in the plane $v_2-\nu^c$ for $\lambda=0.1$}
%    \label{VEVs01}
% \end{center}
%\end{figure}

\begin{figure}[t]
 \begin{center}
\hspace*{1cm}
     \epsfig{file=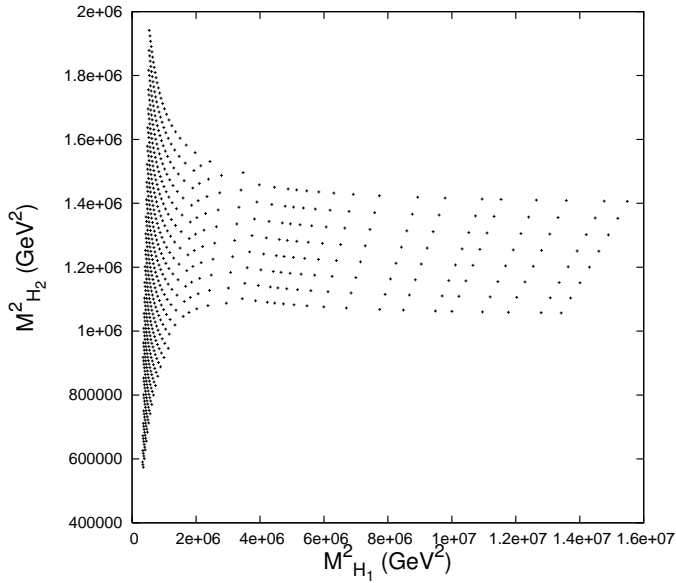 ,height=11.0cm,angle=-90}
\caption{Allowed region by the experimental constraints on the $Z'$ in the plane $M^2_{H_1}-M^2_{H_2}$, for $\lambda=0.1$}
    \label{01masas.eps}
 \end{center}
\end{figure}

\begin{figure}[t]
 \begin{center}
\hspace*{1cm}
     \epsfig{file=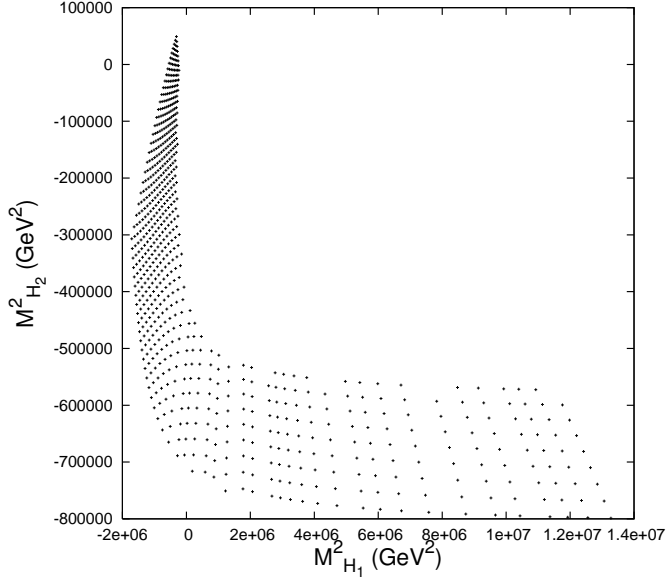 ,height=11.0cm,angle=-90}
\caption{Allowed region by the experimental constraints on the $Z'$ in the plane $M_{H_1}^2-M_{H_2}^2$, for $\lambda=0.3$}
    \label{03masas.eps}
 \end{center}
\end{figure}

%\begin{figure}[htb]
% \begin{center}
%\hspace*{1cm}
%     \epsfig{file=figures/VEVs03.ps ,height=10.0cm,angle=0}
%\caption{Allowed region by the $Z'$ experimental constraints in the plane $v_2-\nu^c$ for $\lambda=0.3$}
%    \label{VEVs03}
% \end{center}
%\end{figure}
One can translate the constraints on the $Z'$ to the plane $(M^2_{H_1}, M^2_{H_2})$, finding the allowed region in the parameter space of the soft masses. We show these regions in Figs. \ref{01masas.eps} and \ref{03masas.eps} for $\lambda=0.1$ and $0.3$, respectively.

Once we have shown that the model 
%We have shown that our model is allowed by the experimental constraints on the existence of a new gauge boson $Z'$. Then, 
is in principle phenomenologically viable, let us now focus our attention on the neutralino sector. 
In the $\mu\nu$SSM with an extra $U(1)$ gauge symmetry, the MSSM neutralinos mix with the extra gaugino. The fact 
that $R$-parity is broken in this model, also produces the mixing of the neutralinos with the left- and right-handed neutrinos. 
%Neutrino masses arise by their mixing with the neutralinos. 
%There are 11 neutralinos in this model. In the approximation of only one family of neutrinos, the neutralino mass matrix would be a $7 \times 7$ matrix. 
Of course, now we have to be sure that one eigenvalue of this matrix is very small, reproducing the experimental results 
about neutrino masses. In the weak interaction basis defined by $\psi ^{0t}=(\tilde Z',\tilde B^0=-i \tilde \lambda ',\tilde W_3^0=-i \tilde \lambda _3,\tilde H_1^0,\tilde H_2^0, \nu ^c, \nu)$, the neutral fermion mass terms in the Lagrangian are  $\mathcal L _{neutral}^{mass}=-\frac{1}{2}(\psi ^0)^t \mathcal M_n \psi ^0+h.c.$, with $\mathcal M_n$ a $7 \times 7$ ($11\times 11$ if we include all generations of neutrinos) matrix,
\begin{eqnarray}
\mathcal M_n= \left( \begin{array}{cc} M & m \\ m^t & 0 \end{array} \right),
\end{eqnarray}
where
\begin{eqnarray}
M=\left( \begin{array}{cccccc}  M'_1 & 0 & 0 & \sqrt{2}g'_1 Q_{H_1} v_1 & \sqrt{2}g'_1 Q_{H_2} v_2 & \sqrt{2}g'_1Q_{\nu ^c}\nu ^c
\\ 0 & M_1 & 0 & -\frac{1}{\sqrt{2}}g_1v_1 & \frac{1}{\sqrt{2}}g_1v_2 & 0 \\ 0 & 0 & M_2 & \frac{1}{\sqrt{2}}g_2 v_1 & - \frac{1}{\sqrt{2}} g_2v_2 & 0 \\ \sqrt{2}g'_1 Q_{H_1} v_1 & -\frac{1}{\sqrt{2}}g_1 v_1 & \frac{1}{\sqrt{2}}g_2v_1 & 0 & -\lambda  \nu ^c & -\lambda v_2 \\ \sqrt{2}g'_1 Q_{H_2} v_2 & \frac{1}{\sqrt{2}}g_1v_2 & -\frac{1}{\sqrt{2}}g_2v_2 & -\lambda \nu^c & 0 & -\lambda v_1+Y_{\nu} \nu \\ \sqrt{2}g'_1 Q_{\nu^c}\nu^c & 0 & 0 & -\lambda v_2 & -\lambda v_1+Y_{\nu} \nu & 0\end{array} \right)\ , \nonumber \\
\label{Matriz Neutralinos}
\end{eqnarray}
is very similar to the neutralino mass matrix of the $U(1)$SSM (substituting $\nu ^c$ by the VEV of a singlet scalar and  neglecting the contributions $Y_\nu \nu$), and
\begin{eqnarray}
m^t=(\ \sqrt{2}g'_1Q_\nu \nu \  \ -\frac{1}{\sqrt{2}}g_1 \nu \  \ \frac{1}{\sqrt{2}}g_2 \nu \  \ 0 \  \ Y_\nu \nu^c \  \ Y_\nu v_2 \ )\ .
\end{eqnarray}
Using typical values of the soft gaugino masses, and with values for the rest of parameters in the region allowed by the constraints on the $Z'$, we have checked numerically that correct neutrino masses can easily be obtained, i.e. once we diagonalize the neutralino mass matrix, one eigenvalue is sufficiently small, of the order of $10^{-2} \ \text{eV}$.
% Then, we can conclude that the neutrino mass generation mechanism works correctly in this model.

However, for the general case of three generations of left- and right-handed neutrinos, unlike what occurs for the 
$\mu\nu$SSM \cite{MuNu SCPV},
the analysis is not so straightforward.
%If we include the three generations in the analysis we can obtain different neutrino mass hierarchies playing with the hierarchies in the Dirac masses. 
% For an extensive analysis of the neutrino sector of the $\mu \nu$SSM, see \cite{MuNu SCPV}. Although such an extensive analysis for the case of the neutrino sector in the $\mu\nu$SSM with an extra $U(1)$ is beyond the scope of this paper, we would like to make a comment on the general case with three generations of left- and right-handed neutrinos.}
As discussed in Section \ref{Section The search of models}, the presence of the extra $U(1)$ group forbids 
the usual Majorana mass term, $\kappa \hat \nu^c \hat \nu^c \hat \nu^c$, of the original $\mu\nu$SSM (\ref{superpotential munuSSM}).
Thus, taking into account the generalization of (\ref{Matriz Neutralinos}) for three generations \cite{Segundo paper munu}, right-handed neutrinos can only acquire large masses through the mixings with the extra gaugino and the Higgsinos due to the terms proportional to $g'_1 \nu_i^c$ and $\lambda^i$, respectively.
Without loss of generality, one can define a basis $\nu_i^c$ such that the VEVs of the right-handed sneutrinos are in the $\nu_1^c$ direction ($\nu_{2,3}^c=0$), while $\lambda^3=0$. Then, only two right-handed neutrinos, $\nu_1^c$ and $\nu_2^c$ can have EW-scale masses, denoted as $M_{\nu_1^c}$ and $M_{\nu_2^c}$. The third one combines with the left-handed neutrinos to form a nearly massles Dirac particle.
As a consequence, the EW-scale see-saw only works for two linear combinations of left-handed neutrinos, that we will denote as $\nu_1$ and $\nu_2$, with a mass given by\footnote{Note that there are other contributions to neutrino masses due to the mixing of left-handed neutrinos with Higgsinos and neutral gauginos \cite{MuNu SCPV}.} $m_{\nu_i} \sim \frac{(Y_{\nu}^{ii} v_2)^2}{M_{\nu_{i}^c}}$ for $i=1,2$, with $Y_{\nu} \sim 10^{-6}$. In general, the four light neutrino states $(\nu_1,\nu_2,\nu_3,\nu_3^c)$ are mixed with the following mass matrix,
\begin{eqnarray}
%M_{light}=
\left( \begin{array}{cccc} m_{\nu_1} & 0 & 0 & m_D^1 \\ 0 & m_{\nu_2} & 0 & m_D^2 \\ 0 & 0 & 0 & m_D^3 \\ m_D^1 & m_D^2 & m_D^3 & 0 \end{array} \right)\ , 
%\nonumber \\
\label{M light}
\end{eqnarray}
where the state $\nu_3$ is orthogonal to $\nu_{1,2}$ states and $m_D^k \simeq Y_\nu^{k3} v_2$, with $k=1,2,3$, are the Dirac masses in this basis. At tree-level, there are four light Majorana states. To account for neutrino data, for example the cosmological bound from WMAP on the sum of all light neutrino masses, of the order of the eV, a fine-tuning on the other three entries of the Yukawa matrix is necessary, $Y_\nu^{k3} \leq 10^{-11}$.
% with $k=1,2,3$ and $l=3$.
We have checked this result numerically. For a more detailed discussion of this mechanism see \cite{B-L models}, where a similar situation occurs in the context of $U(1)_{B-L}$ supersymmetric models with broken R-parity, although in that case the EW-scale see-saw works only for one neutrino.

Therefore, the $\mu\nu$SSM with an extra U(1), in the general case of three generations of neutrinos, predicts the existence of two heavy right-handed neutrinos, of the order of the TeV, and four light (three active and one sterile) neutrinos. 
Given the oscillation anomalies in LSND, MiniBooNE and MINOS \cite{LSND,MiniBooNE,Minos}, 
the extra light sterile neutrino might be welcome \cite{Akhmedov}.

Finally, we have also performed an estimation of the tree-level upper bound on the lightest Higgs mass in this model. Let us recall that, neglecting the small neutrino Yukawa coupling effects, the expression of the upper bound on the lightest Higgs mass in the $\mu \nu$SSM is equivalent to that of the NMSSM, once we define $\lambda^2=\lambda_{1}^2+\lambda_{2}^2+\lambda_{3}^2$, and this is given by the following tree-level expression \cite{Segundo paper munu}:
\begin{eqnarray}
m_h^2 \leq M_Z^2 (\cos^2 2 \beta + \frac{2 \lambda^2 \cos^2 \theta_W}{g_2^2}\sin^2 2\beta)\ .
\end{eqnarray}

This bound receives a positive contribution from the extra $U(1)$ sector \cite{Contribucion U1 a cota masa Higgs mas ligero},
%: $2 g'_1 v^2(Q_{H_2} \cos^2 \beta+ Q_{H_1} \sin^2 \beta)^2$. 
in such a way 
%that Adding the two contributions we conclude
that the tree-level formula for the upper bound on the lightest Higgs mass in the $\mu \nu$SSM with an extra $U(1)$ is given by:
\begin{eqnarray}
m_h^2 \leq M_Z^2 (\cos^2 2 \beta + \frac{2 \lambda^2 \cos^2 \theta_W}{g_2^2}\sin^2 2\beta)+2 g'_1 v^2(Q_{H_2} \cos^2 \beta+ Q_{H_1} \sin^2 \beta)^2\ .
\label{upper new}
\end{eqnarray}

It is worth recalling here that the LEP lower bound on the Higgs mass is $114$ GeV. In the case of the MSSM, the tree-level upper bound on the lightest Higgs mass is $M_Z$, i.e. smaller than $114$ GeV. As is well known, the MSSM is not still ruled out by LEP because the radiative corrections can raise this upper bound, although this can require some fine-tuning. In the case of the NMSSM, once perturbativity is imposed, the tree-level upper bound on the lightest Higgs mass can be as high as $110$ GeV, improving the Higgs-mass problem of the MSSM. In \cite{Segundo paper munu}, this issue has been analyzed in the context of the $\mu\nu$SSM, and the upper bound turns out to be similar to the one of the NMSSM.

In the $\mu\nu$SSM with an extra $U(1)$, the upper bound depends on the value of 
the extra gauge coupling $g'_1$, as shown in (\ref{upper new}). Whereas for $g'_1 \simeq g_1$ this bound 
is only raised to $113$ GeV, 
%We have estimated that for reasonable values of this coupling, 
for $g'_1 \simeq 2 g_1$ it is raised to about $120$ GeV.
%\textbf{while for $g'_1 \simeq g_1$ it is only raised to $113$ GeV}. 
Thus, the addition of an extra $U(1)$ gauge group to the $\mu\nu$SSM has also the nice feature of increasing the tree-level bound on the Higgs mass leading to a larger window for the discovery of the Higgs at collider experiments.
%, in agreement with the recent hints of a Higgs boson with a mass about $140 \ \text{GeV}$ at LHC.

\section{Conclusions}\label{Section Conclusions}
% In this paper, we have studied an extension with an extra $U(1)$ gauge symmetry of a supersymmetric model which solves the $\mu$ problem at the same time that generates the neutrino masses: the $\mu\nu$SSM. The superpotential of this model includes R-parity violating terms and, just like the NMSSM, presents a cosmological domain wall problem. We have used the extra gauge symmetry to forbid dimension 4 and dimension 5  B-violating operators to ensure the stability of the proton and, at the same time, to solve the cosmological domain wall problem. We have searched for consistent models using the anomaly cancellation conditions and selecting which terms are allowed in the superpotential. Exotic matter should be added to the spectrum to cancel all the anomalies. This extra matter should be sufficiently massive to have escaped the detection. We have included effective mass terms for the exotic matter in the superpotential once the gauge singlet sneutrinos take VEVs in the EW breaking.

The $\mu\nu$SSM solves the $\mu$ problem of the MSSM and generates correct neutrino masses by simply using 
right-handed neutrinos. This mechanism implies that only dimensionless trilinear terms, breaking $R$-parity, are present in the superpotential.
The non-presence in the superpotential of proton decay operators breaking $R$-parity, a trilinear term generating a domain wall problem, and bilinear terms such as the $\mu$ term and the Majorana masses, is solved in the
$\mu\nu$SSM using string theory arguments, discrete symmetries or non-renormalizable operators.
In this work we have used a different strategy, namely an extra $U(1)$ gauge symmetry is added to the gauge group of the Standard Model. Since all fields of the $\mu\nu$SSM can be charged under the extra $U(1)$, all the dangerous operators mentioned above could in principle be forbidden. We have checked that this is precisely the case. For example,
dimension four and five baryon number violating operators are forbidden in the superpotential, ensuring
the stability of the proton. 

On the other hand, the anomaly cancellation conditions associated to the extra $U(1)$, allow us to constrain the values
of the $U(1)$ charges. We have checked that six assignments of the $U(1)$ charges to the matter fields
are viable, once extra matter is introduced.
In particular, three generations of vector-like color triplets and $SU(2)_L$ doublets, as well as six Standard Model singlets are needed. 

We have studied the phenomenology of the model focusing our attention on the electroweak symmetry breaking.
We have found that it is viable, with wide regions of the parameter space fulfilling the experimental constraints on the existence of a new gauge boson $Z'$.
We have also studied the neutralino sector of the model, since the neutrinos and the extra gaugino mix with the MSSM neutralinos. We have checked numerically that the experimental results on neutrino masses can be reproduced.
Nevertheless, the analysis of the generalized see-saw matrix in the case of three generations is different from the usual one in the $\mu\nu$SSM. This is because of the absence of the trilinear term generating the domain wall problem, but also an effective Majorana mass term. Due to the absence of the latter, the EW-scale see-saw works only for two neutrinos and therefore one needs some entries of the Yukawa matrix small.
Finally, we have estimated the tree-level upper bound on the lightest Higgs mass, finding that it can be as large as about 120 GeV. \\

\noindent {\bf Acknowledgments} 

\noindent This work
was supported in part by the MICINN under grants FPA2009-08958 and FPA2009-09017, 
by the Comunidad de Madrid under grant HEPHACOS S2009/ESP-1473, and by the
European Union under the Marie Curie-ITN program PITN-GA-2009-237920.
The authors also acknowledge the support of the Consolider-Ingenio 2010 
Programme under grant MultiDark CSD2009-00064.

\end{document}